\documentstyle[12pt,worldsci]{article}
\newcommand{\be}{\begin{eqnarray}}
\newcommand{\ee}{\end{eqnarray}}
\def\lo{\langle 0 |}

\def\fc{ f_{\eta'}^{(c)} }
\def\gmf{\gamma _{5}}

\def\ra{ \rangle }

\def\gmmu{\gamma _{\mu}}

\def\gmf{\gamma _{5}}

\begin{document}
\title{(PARTIAL) SUMMARY OF SECTION A: \\ CONFINEMENT MECHANISM, FLUX
  TUBES AND STRINGS  }
\author{Edward Shuryak\\
{\em Department of Physics and Astronomy \\
 State University of New York, Stony Brook Ny 11794, USA}\\
%\vspace{0.3cm}
%and \\
%\vspace{0.3cm}
%SECOND AUTHOR'S NAME\\
%{\em Group, Company, Address, City, State ZIP/Zone, Country}}i
}
\maketitle
\setlength{\baselineskip}{2.6ex}

\vspace{0.7cm}
\begin{abstract}
 An overview of some issues discussed is given, including the
 following ones: (i) the mechanism of confinement, (ii) the structure of
 the string, (iii)  non-perturbative glue in vacuum and hadrons,
 and (v)  quark-hadron duality. In doing so I try 
to indicate issues important for JLab.
Finally I discuss
  another issue debated during
this conference, namely (vi) 
whether one can use some common effective potential for
description of various hadrons, using heavy quarkonia as an example.

\end{abstract}
\vspace{0.7cm}

\section{Overview}
 My first remark is about a general situation in non-perturbative QCD/
hadronic physics. It is striking that it has several sub-communities,
with rather weak interaction between them.
Roughly speaking, there are three different schools which think 
about those issues in completely
different terms. 

(i) The first one (naturally, the dominant one at this conference),
concentrate on {\it confinement} phenomenon and strings. 
Hadrons are viewed basically as bare
quarks connected by strings, and hybrids are string
excitations. The vacuum is ``empty'' of any
non-perturbative glue and the glueballs are
closed strings.
 The stronghold of this point of view is physics of charmonium/bottonium and
Regge
trajectories. There are of course
extended claims 
for  all other hadrons as well. (These proceedings is probably
the best possible review of development in these directions.)
  
(ii) The second one (to which I belong) concentrate at hadrons made of
light quark. The basic phenomenon underlying it is {\it
breaking of  chiral
  symmetry }, both spontaneous ($SU(N_f)_A$) and explicit by the anomaly ($U(1)_A$). This school view light quarks as
collective excitations propagating on top of the quark condensate 
$<\bar q q >$, and thus obtaining ``constituent quark" mass, being
 the main part of masses of all light hadrons. The residual interaction
binds those into hadrons, in some cases even without a confining potential
(which is treated as small effect). 

Originally this school was using Nambu-Jona-Lasinio-type models. 
Recent QCD-based development views
  the vacuum (as well as  glue inside glueballs and
 hadrons) as a superposition of
   {\it instantons}, which naturally provide the (non-local)
interaction between quarks needed for  chiral breaking.
 The obvious stronghold of this
 ideology  is chiral dynamics of pions and its relatives. There is
 also a claim that all other lowest mass hadrons made of light quarks
(and heavy-light ones) 
are explained by it. For recent review I propose that by T.Schaefer
and myself \cite{SS_98}.
  
 (iii) The third school tries to {\it re-sum perturbative diagrams} using
 well
known QED tools like Schwinger-Dyson or Bethe-Salpeter equations. They
hope that when the coupling constant is not too small, both chiral
  symmetry breaking and confinement will appear, e.g. from  modified
  propagators. An example is the $G\sim 1/q^4$ behavior of the gluon
propagator at small q. Similarly, the one-gluon exchange, if strong enough,
 may also lead to
chiral symmetry breaking,
 quark constituent masses etc.

  The first point I want to make is that these three pictures of
  hadrons are indeed so different, that they cannot be simultaneously
  right. And available lattice/phenomenological data should be
used to pick the right one. 
One example (which was several times mentioned at this conference):
 although the
  $G\sim  1/q^4$ gluon propagator may provide confinement, it definitely
does $not$ correspond to a string-like
  configuration of the field. Thus  it is already ruled out by
  lattice studies, which do see a rather narrow
 string between  static charges. (Other similar examples are to follow below.)

My next general point provides an argument
$against$ the perturbative views (iii).
In fact we do not have $one$ QCD
vacuum / hadronic spectrum but $many$,  numbered by the famous $\theta$ angle
related with the phase of the instanton amplitude. For example, at $\theta=0$
   (the physical case) and  $\theta=\pi$ there is no CP violation, but
  the  instanton-induced
amplitude change sign.
Strong arguments exist showing that there should be
 drastic
differences between these vacua, and as $\theta$ changes
 different vacua cross should each other and cause some phase transitions.

The perturbation theory (no matter how resummed, with or without renormalons)
simply does not know
$anything$
about the $\theta$ angle! What it means is that there cannot be any unique
way of defining QCD starting from perturbative series, unless it somehow
includes this parameter.

The same question applies to   ``abelian projected'' version of QCD
(much advocated at this conference). It is not a trivial question
whether it may or may not generate
the topologycal charge, 
  to know about the $\theta$ angle as well.

  Of course this is not a new argument. Still it is worth repeating
  that $\theta$ dependence is not an academic issue. In fact
dependence
is very strong.   For example,  in gauge theory without quarks
one can se directly from the lattice data on topological susceptibility that
  the energy difference between the two vacua mentioned
above, $\theta=0,\pi$ is about a 1 GeV/fm$^3$, larger
 than energy densities associated with chiral breaking or
 confinement
in all models. And it would be difficult to believe that any model
which completely misses 
such large effects can be successful for description of smaller ones.

\section{The Physical Mechanism of Confinement}

  Impressive lattice studies reported 
(see especially talks by Bali and Michael) 
has strengthened the case made by earlier works
 on the field distributions in static configurations.
Properties of glue excitations with different orbital momenta they found
 leave little
doubt that a (rather thin) string is indeed formed in such configurations.

 This by itself rules
out some models of confinement. I already mentioned the model with the
gluon  propagator  $G\sim 1/q^4$. Another one which I think belongs to the same class is a suggestion \cite{large_inst}   that confinement is due
to the tail of the instanton distribution over sizes $dn/d\rho$ at large 
$\rho$: I will return to it at the end of this section. 

  {\it What is the internal structure of a string?} 
It seems that lattice data are consistent with the dual
  superconductor picture, see talks by  Bali, Polikarpov, Ichie and Suganuma.
   In addition, as M.Baker told us, the dual picture
   reproduces the spin-dependent potentials well. 

The interesting discussion  here was initiated by the observation that abelian
projected field shows string radius to be only about .15 fm, while the
complete SU(2) fields indicate about twice larger radius. The point
(emphasized  by Suganuma, but also by others) is that the non-abelian
fields have to be
located outside the abelian one, since the monopole current 
should wind around abelian field,  and 
the monopoles are made of it.

 This convincing argument however  seem somewhat to
contradict another set of ideas, presented here by Kondo and  Suganuma.
They proposed that the non-abelian fields can be to first
approximation
ignored, and then taken into account perturbatively. The reason for
that,
they argued, is because its phase
is a simple
``random variable''. However it seems to me
 that this view point cannot be quite correct, since it would destroy the
monopoles and evantuallu the outer
region of the string.

 {\it How the string interact?} Recall that attraction means
dual superconductor of the 1-st kind, and repulsion means the 2-nd.
Bali has shown that there are clear signals of weak attraction.
C.Michael has also studied the issue, and his conclusion is the 
octet string tension is nearly exactly twice the triplet
one\footnote{The famous ``Casimir scaling'' for the non-triplet
  strings was debated in literature for ages: it seem not to work.
I never understood what
it is based on anyway.}: thus he sees no
significant interaction. It also suggests that QCD
strings are close to the border line.

   {\it The theoretical status of the dual superconductor picture} is
   however still
   controversial.  Its proponents say that recent
Seiberg-Witten solution of N=2 Supersymmetric QCD 
 has provided support to their point of view.
Well, not quite. In this theory the  Higgs VEV does break
the gauge group, and 
both monopoles and
dual photons are indeed real excitations of the theory. In QCD we definitely
 have no
``dual gluons'' or Higgses with masses $\sim 1 GeV$ \footnote{The data for
  shown by Suganuma indicate that the abelian part
has the propagator which is consistent  with $zero$ mass. Has it
any specific
physics consequences?  }. The fact that those do appear
in the abelian dual Landau-Ginzburg Lagrangian basically invalidate
it,
no matter how good is its description of the string. It is not the right
effective theory of QCD at low energies: we have to look for 
another one, without the color
group
being broken.

{\it Is confinement generated by monopoles?} It was repeatedly
mentioned
that several abelian projections were shown to preserve the full
string tension. Another strong contender
   (discussed here by Polykarpov)
 is further reduction of the abelian fields to only $Z_N$ parts of
 them.
As shown in \cite{Greensite} it also  reproduces the string
tension, reviving an old viewpoint that
 $vortices$ rather then  monopoles (or the 2-d objects rather
 than the 3-d ones) are responsible for confinement.
A debate which of the two are more physical objects is 
still going on, and extensive lattice work is needed to resolve the issue.

  Let me finish this section with a discussion of two (instanton-related)
confinement mechanisms
which (I think) do $not$ work. The first is based on the correct
observations that
(i) an abelian projection of an instanton produces a monopole loop (of
size $\rho$) and  (ii) in a sufficiently dense  ensemble of instantons
those
loops merge into a long loops, in a manner similar to a percolation
transition. Nevertheless, the potential V(r) (extracted by calculating the
Wilson loops with the ensemble of instantons) show smooth (and
approximately linear) dependence on the instanton density, ignoring
the percolation point.  

The second is the suggestion \cite{large_inst} that large-$\rho$ tail of the
instanton distribution may be responsible for confinement.
Indeed, if $dn/d\rho \sim \rho^{-3}$ one can get it. However, if the
origin of confinement are kind of large domains of weak field, 
we would not get thin strings. Furthermore, if it is due to some tail
of the distribution, confinement should be ``intermittent'' on
configuration-by-configuration basis: many lattices should have no
confinement and some have very strong one. It is completely contrary
to lattice observations, in which confinement fluctuates little.

With a tail $dn/d\rho \sim \rho^{-5}$ which fits well
 (still rough) lattice data, one gets approximately linear
potential $V(r)=Kr$
up to r$\sim$ 3 fm. There was a claim by Suganuma et al
\cite{suganuma} that the tension $K$ has the right
magnitude, but it must be mistaken since our calculation \cite{CNS_98} as
well as by others \cite{Boulder,pot_others} get in this case the tension
$K\sim 100-200 MeV/fm$, few times less than the empirical/lattice
value $K\sim 1 GeV/fm$.

\section{The Non-perturbative Glue}
\subsection{ How the Non-perturbative Glue in Vacuum  actually look like?}
Some people reject  this question\footnote{ Paton commented for
example that in his model the vacuum has no non-perturbative glue at
all.},
but I think it remains to be the most important issue, a key to
 the whole hadronic physics. 
Let me make the question more specific, emphasizing dimensionality
of the objects:
 {\it Is it mostly made of (i) closed string (2-d membranes), (ii)  instantons (4-d objects),
  (iii)  monopoles (3-d objects) or (iv)  vortices(2-d objects)?}

Obviously our lattice friends can  answer this question
simply by making lattice fields smooth (in order to get rid of
``quantum noise'' of perturbative nature) and then see what they look like.
What they see are 4-d ``bumps''. Many of them have nonzero topological
charge and are identified as instantons or anti-instantons. Some have zero 
topological charge and smaller action, they look like close 
instanton-anti-instanton pairs (and  disappears under further
``cooling'', oraction
minimization). Something of the order of 90 percent
 of the action is saturated by those objects. 

For those who  think that a ``smoothening'' of gauge fields is 
a questionable procedure, there is another way of looking at the problem, this time from
the point of view of light quarks. One can look how the wave function
of the quark condensate (which all approaches mentioned above claim to
generate). In order to do so, one should
take $|\psi_\lambda(x)|^2$ for the lowest eigenmodes $\lambda$
of the Dirac operator and study how the corresponding eigenfunction is
distributed in space. What one sees on the lattice are again
isolated bumps correlated to  instantons.

 I never head that
 anybody was able to
identify 3-d (or 2-d) structures either in action distribution
or in quark condenate: for me it
put large question mark for claims that monopoles (vortices) are physical. 
The ``instanton dominance'' is clealry established, as far
as
gauge field action of chiral symmetry breaking is concerned. However
it is not so if we for  confinement (see the previous section).

  In fact, a ``slightly smoothened'' vacuum preserves confinement
  with
the original string tension (see e.g. \cite{Ilgen}), and further
studies should be able to find out what it is due to.

\subsection{Glue in Glueballs and Hybrids}

At this conference this subject was discussed from a string picture
point of view. It does not work for recently claimed hybrids,
as those turned out to be lighter than expected.
Let me present an argument that it cannot be true at least for the scalar
glueball
as well, based on its size and shape.

 The $size$ of the scalar glueball
 is surprisingly small.      This was first seen on the lattice
from the magnitude of finite size effects
      \cite{BK_89} and then confirmed  directly from measurements of the wave
      functions \cite{ISST_83,FL_92}. The size of the scalar glueball
      (defined through the exponential decay of the wave function)
      is $r_{0^{++}}\simeq 0.2$ fm, while $r_{2^{++}}\simeq 0.8$ fm
      \cite{FL_92}\footnote{
For comparison, a similar measurement for the
$\pi$ and $\rho$ mesons gives 0.32 fm and 0.45 fm.}. It indicates
that spin-dependent forces between gluons are much stronger
than between quarks, at least in the scalar channel.
Instanton-based calculation done by T.Schaefer and myself \cite{SS_glueballs}
have obtained both the mass and size values of the scalar glueball
in agreement with those obtained on the
lattice. 
The
parameter enhancing gluon interaction with instantons is nothing else
but large classical amplitude of the instanton  field at the center  
$G^2(0)=192/(g^2\rho^4)$ (basically, a parameter
 of the same nature as the one enhancing
induced photon emission inside lasers). 

 The wave function of the
scalar glueball
(defined by a split-point correlator) has a strong peak at the
origin. 
This is in direct contradiction with the closed string picture,
suggesting a hole there.  

Closing this subsection, let me say that all these arguments do not of course
invalidate estimates based on the string picture: it is  possible
(and even quite probable)
that resonant
excitations of many different types can coexist, as they do in countless
examples from condense matter/nuclear physics.

\subsection{How to look for Glue inside Ordinary Hadrons?}

One method,  is to extract
the {\it next twist matrix elements} for nucleons from power
corrections to deep inelastic scattering.
The theory of these corrections  based on general Operator Product Expansion
(OPE)  was worked out in early 80's \cite{highertwists}. As emphasized
in these works, while the leading twist operators are related to
$probabilities$ to
find quarks or gluons with certain momenta inside the nucleon, the
next twist ones describe $correlations$ between partons. In
particularly,
mixed quark-gluon operators of the structure $<N|\bar q ... G_{\mu\nu}... q |N>$ 
tell us what is the gluon field at the same (scattering) point where
the quark is. Needless to say, it would be very important to have such
information, and JLab experiments are well suited to fill this gap.
   
  Meanwhile, one can estimate such matrix elements from various
  models of the nucleon. Such estimates based on the instanton liquid
  model \cite{polyakov} were
discussed here by M.Polyakov. His results show that in fact these
matrix elements are smaller than a naive estimates based on
the magnitude of the fields inside the instantons: estimates have some 
small factors like diluteness of the instanton vacuum. The reason for
that
is that the quark operators which appear as corrections to DIS all are
vector or axial vectors, so they are not chirality changing and cannot
thus be represented by instanton zero modes. It is a part of a general
tendency of the instanton effects to be ``hidden'' in vector and
axial channels. 

  Much larger gluonic effects are expected in scalar/pseudoscalar
  channels or isosinglet axial channel related with the anomaly.
Very intriguing tool in this respect is 
{\it  virtual/real charm annihilation} processes. Let me conclude this
section with three examples of this kind.

(i) The process $\eta_c \rightarrow \bar N N$ is forbidden by
chirality conservation rules of perturbation theory: nevertheless
it is observed and has a surprisingly large branching. Its relation to
instantons was discussed in \cite{AF}.

(ii)Recently, CLEO collaboration has reported 
 measurements of inclusive and exclusive production of 
the $ \eta' $ in B-decays :
\be
\label{1}
Br( B \rightarrow \eta' + X \; ; 2.2 \; GeV < E_{\eta'} <
2.7 \; GeV ) = \nonumber \\ (7.5 \pm 1.5 \pm 1.1) \cdot 10^{-4} \; ,  
\ee
\be
\label{2}
Br(B \rightarrow  \eta'+ K ) = (7.8_{-2.2}^{+
2.7} \pm 1.0) \cdot 10^{-5} \; .
\ee
Simple 
estimates  show that these data are in  severe 
contradiction 
with the standard
 $b$-quark decay into 
 light quarks:
Cabbibo  suppression
$V_{ub}$ leads to
decay rates
two orders of magnitude  smaller than the data (both  
inclusive and exclusive ones). 
    Alternative mechanism, suggested
by Halperin and Zhitnitsky \cite{HZ},
is based on the Cabbibo favored
$ b \rightarrow c \bar{c} s $ process, followed by a transition
of virtual $ \bar{c} c $ into the $ \eta' $. The latter transition implies
large ``intrinsic charm'' component of the $ \eta' $. Its  
 quantitative measure can be expressed through the matrix element
\be
\label{3}
\lo \bar{c} \gmmu \gmf c | \eta'(q) \ra \equiv i \fc q_{\mu} \; . 
\ee
and one needs
$\fc \approx 140 \; MeV$  
in order to explain the CLEO data, see \cite{HZ}.
 This value is surprisingly large, being 
 only a few times smaller than the 
analogously normalized residue $ \lo \bar{c} \gmmu \gmf c 
| \eta_c (q) \ra = i f_{\eta_c} q_{\mu} $ with $ f_{\eta_c} 
\simeq 400 \; MeV $ known experimentally from the $ \eta_c 
\rightarrow \gamma \gamma $ decay. 

 By expansion in inverse powers of charm quark mass, the problem can be reduced to the 
matrix element of a particular dimension-6 pseudo-scalar gluonic operator:
$
\lo g^3 f^{abc} G_{\mu \nu}^a \tilde{G}_{\nu \alpha}^b 
G_{\alpha \mu}^c |\eta'>   
$.
It was estimated in \cite{SZ} and found to be indeed large, 
enhanced by very strong gluonic 
fields of small-size instantons. Quantitative description of CLEO data
still remains to be work out, however.

(iii) The last example (strongly related with the previous one) is the
contribution to the spin of the nucleon of  the polarized charm ``sea''
(to be measured at next generation DIS experiments like CERN COMPASS).
It is also related with the matrix element of the current $\bar c
\gamma_\mu \gamma_5 c$, but this time averaged over the
nucleon. As explained in recent paper by Blotz and myself \cite{BS},
using the same gluon operator one can do it
 in the instanton model. The resulting charm contribution of about -0.03.

\section{Quark-Hadron Duality}

The issue in early 70's was a check of QCD itself: and we did found
that,
for example, the total cross section of $e^+e^-\rightarrow hadrons$
is close to total cross section of annihilation to all species of quarks. 
Now the topic was revived, and the issue is the {\it 
nature and magnitude of the oscillating
component}, the difference between the two.  

 One useful direction of recent work is to study what happens in
exactly solvable models.
At this conference Lebed has presented results \cite{GL} for heavy
quark decay in 2-d QCD at large N,
known as 't Hooft model. The issue was also studied by
Shifman et al  \cite{Shifman_etal}. Two groups have nicely
demonstrated that the duality itself works. However they strongly
disagree about the magnitude of oscillating component. Lebed claimed
that it is O(1/M), based on numerical fit,
 while Shifman et al find much smaller effect
O(1/M$^3$)
(M is the mass of the heavy quark).

 What is the general reason why the oscillating part exist at all?
Shifman et al \cite{Shifman_etal} simply argued that since the exponentially decaying
contribution to correlators for space-like
(Euclidean) momenta $\Pi(Q)\sim O(exp(-Q*const))$ does exist ,  such contributions become
oscillating O(sin(Q*const)) for the time-like (Minkowski) momenta.
What I find exciting about this argument is that oscillation period is
related with 
singularities of amplitudes in x space away of x=0 (like the radius of
the instanton in expressions like $1/(x^2+\rho^2)$).

Let me add two important comments on that. First, since the derivative of
the scattering
amplitude over  energy  determines the {\it lifetime of the
  intermediate system}, the oscillating component should exist because
otherwise we would not be able to get from the amplitude the duration of
the final sate interaction. (In the canonical case of
total cross section of $e^+e^-\rightarrow hadrons$, this is the time
at which the string is broken and final hadrons appear, or widths
of corresponding resonances.) 

The second: one can ask what happens when resonances overlap\footnote{For
  example, in Regge trajectories  $M_n \sim \sqrt(n)$ and spacing
  between resonances is $O(1/ \sqrt(n)$ while the width $\Gamma_n\sim
  M_n$.}. My general expectation is that oscillating function is
changed to a $random$ one. However its correlators like
\be K(\epsilon, \delta\epsilon)=<\sigma(\epsilon+ \delta\epsilon) \sigma(\epsilon)> \ee
should be regular functions which
 still display the characteristic
correlation scale in energy of the order of the inverse time of
final state interaction $\delta\epsilon\sim
\Gamma(\epsilon)$.
Similar phenomenon known in nuclear physics under the name of
Erickson fluctuations.

Summary: the physical nature and accuracy of parton-hadron duality is
not yet clarified. Clearly much more theoretical and experimental work
is needed. A good place for the latter is 
 Jefferson Lab. which can check on accuracy of/deviations from the so called
Bloom-Gilman
duality (a difference between  sum over  hadronic excitations
and  smooth  partonic structure function).

\section{Can one use a common Effective Potential for various hadrons?}

  This issue was not part of section A I have to review, but it was
  discussed in a very interesting lecture by P.Lepage and, since it is
  vital
to the field, I am tempting to try to clarify the issue somewhat.

Let me briefly repeat what he said in his talk\footnote{
The reader of the proceedings can of course read the Lepage's
contribution, 
while now I only have recollections of his talk. In case my points are
also made there, I apologize.}. His basic idea is that lattice
simulation allows
one to get wave functions of quarkonia states like $J/\psi$ and
$\Upsilon$,
which can in turn be used to get an effective potentials (by putting
those
into Schroedinger equation). Among the questions Lepage has addressed
are: (i) whether these
potentials are universal and (ii) how close they are to the one
extracted for static quarks. His results show that (i) for  $J/\psi$ and
$\Upsilon$ the potentials are very close, but (ii) both are about 20 percent
different from the static one.

   Let me start with the issue of the time scales relevant for the
   problem. We discuss heavy quarks Q interacting with a
   non-perturbative glue,
so let us imagine that those have characteristic time scales $\tau_Q$
and $\tau_g$ respectively.

  The static potential would be correct if $\tau_Q >> \tau_g$. It is
  however easy to see that the heavy quark limit ($M\rightarrow \infty$)
in fact leads to the $opposite$ relation. It is true that the velocity
of heavy quarks decreases with M, but sizes of quarkonia decrease even more 
and $\tau_Q\sim 1/(M \alpha_s^2)$. From old papers of Voloshin and
Leutwyler  we know that one can use in this limit dipole
expansion,
leading to oscillatory $V\sim r^2$ potential with the 
state-dependent coefficient. This explains why Lepage's upsilon potential
deviates from the static one.

But why it is the same for $\Upsilon$ as for  $J/\psi$? My answer to that is that (if
we still use the dipole approximation to second order) the effective
potential
is proportional to the following vacuum average over the gluoelectric fields
$$ V \sim r^2 \int d\tau <\vec E(\tau)U \vec E(0)U^+> exp(-\tau \omega_{PS}) $$
where U are color trasnport matrices and 
$\omega_{PS}$ is the energy difference between the original S-wave
state   ($J/\psi$ or
$\Upsilon$) and that of P-wave states the dipole excites\footnote{For
  simplicity
I ignore their splittings.}. 

Although the charmonium states are mostly sensitive to
 the linear potential while
bottonium are more due to the Coulomb part, by some play of numbers the
essential frequency $\omega_{PS}$ is
about the same for both. It means that the frequency of rotation of
quarks
in both states are the same, and we probe the $same$ correlator of the vacuum 
fields. (Incidentally, it is  about coincides with the correlation
time of the vacuum field, see e.g. the talk by Dosch in these proceedings.). 

Of course, if we go outside the dipole approximation in the second
order,
there are differences between two cases (e.g. the relativistic
corrections with higher power of the velocity).  Lepage's results then
imply that those are sufficiently small.

In summary: neither very heavy quarkonia (nor real $J/\psi,\Upsilon$)
should be described by static potentials. One may construct those
using lattice data for wave functions (and/or Gauge field
correlators), but those are going to be state-dependent.

%\section{Summary and Discussion}

\end{document}